\documentstyle[12pt]{article}

\newcommand{\hh}{\rule[-5mm]{0mm}{10mm}}

\newcommand{\bz}{{\hbox{\bf Z}}}
\newcommand{\br}{{\hbox{\bf R}}}
\newcommand{\bc}{{\hbox{\bf C}}}

\newcommand \qint[1]{\left[ {#1} \right]}
\newcommand \bosal[5]{\, a \! \left(
{#1};{#2},{#3} \, \vrule \, {#4} ; {\textstyle {#5}} \right)\,}
\newcommand \bosbl[5]{\, b \! \left(
{#1};{#2},{#3} \, \vrule \, {#4} ; {\textstyle {#5}} \right)\,}
\newcommand \boscl[5]{\, c \! \left(
{#1};{#2},{#3} \, \vrule \, {#4} ; {\textstyle {#5}} \right)\,}
\newcommand \bosa[3]{\, a \! \left(
{#1} \, \vrule \, {#2} ; {\textstyle {#3}} \right)\,}
\newcommand \bosb[3]{\, b \! \left(
{#1} \, \vrule \, {#2} ; {\textstyle {#3}} \right)\,}
\newcommand \bosc[3]{\, c \! \left(
{#1} \, \vrule \, {#2} ; {\textstyle {#3}} \right)\,}
\newcommand \diff[2]{{~}_{\scriptstyle {#1}}
\displaystyle \partial_{\scriptstyle {#2}} \,}
\newcommand \fra[2]{\displaystyle
{\frac{\textstyle {#1}}{\textstyle {#2}}}}

\begin{document}

\begin{flushright}
  UT-617 \\
  (Revised Version) \\
  Sept. 1992
\end{flushright}
\vspace{24pt}
\begin{center}
\begin{large}
{\bf Free Boson Representation of $U_{q}(\hat{sl_{2}})$}
\end{large}

\vspace{36pt}
Jun'ichi Shiraishi

\vspace{6pt}

{\it Department of physics, University of Tokyo }\\
{\it Bunkyo-ku, Tokyo 113, Japan }

\vspace{48pt}

\underline{ABSTRACT}

\end{center}

\vspace{4cm}

A representation of the quantum
affine algebra $U_{q}(\hat{sl_{2}})$ of an arbitrary
level $k$ is realized in terms of three boson fields, whose $q \rightarrow 1$
limit becomes the Wakimoto representation. An analogue of the screening
current is also obtained. It commutes with the action of
$U_{q}(\hat{sl_{2}})$ modulo total difference of some fields.

\vfill
\newpage

{\bf 1. Introduction} \qquad Recently the anti-ferroelectric spin 1/2
XXZ-Hamiltonian was exactly diagonalized \cite{DFJMN} by
using the technique of $q$-vertex operators \cite{FR}.
(See \cite{IIJMNT} for higher spin cases.)
Further, in \cite{JMMN}, an integral
formula for correlation functions of local operators was found
in the case of spin 1/2 with the help of a boson
representation of $U_{q}(\hat{sl_{2}})$ of level $1$ \cite{FJ}.
However, for the higher spin XXZ model,
it seems difficult to get such a formula because we lack
the boson representation of $U_{q}(\hat{sl_{2}})$ for higher level.
The aim of this paper is to show that $U_{q}(\hat{sl_{2}})$ is also
bosonized for an arbitrary level.
As in \cite{FJ}, we shall use an analogue of the currents defined by the
Drinfeld realization \cite{Dr} of $U_{q}(\hat{sl_{2}})$ and express them by
three boson fields.
In the $q \rightarrow 1$ limit,
this new representation tends to the Wakimoto
representation with bosonized $\beta - \gamma$ system \cite{Wa}\cite{FMS}.

We obtain an analogue of
the screening current in terms of boson fields. This operator has
the property that it commutes with the currents modulo total
differences of some operators.
Hence a suitable Jackson integral of the screening current should
commute exactly with the currents.
A Jackson integral formula for the solution to
$q$-deformed Knizhnik-Zamolodchikov equation \cite{FR}
was found by
Matsuo \cite{Ma1}\cite{Ma2}
and Reshetikhin \cite{Re}. We think that there is a deep connection between
the existence of our screening current and these Jackson integral formulas.

After finishing the work we learned from Atsushi Matsuo that he
got another bosonization \cite{Ma3}.
The relation between the two bosonizations is yet
to be clarified.

\vspace{1cm}


{\bf 2.  Free boson fields $a, b,$ and $c$} \qquad
In this article we consider bosonization of the Drinfeld realization of
$U_{q}(\hat{sl}_{2})$.
We construct Drinfeld's generators in terms of three free boson fields.
Hereafter let $q$ be a generic complex number
such that
$|q|<1$. We will frequently use the following standard notation:
\begin{equation}
\qint{m} = \fra{q^{m}-q^{-m}}{q-q^{-1}},
\end{equation}
for $m \in \bz$.

Let $k$ be a complex number.
Let $\{a_{n},b_{n},c_{n},Q_{a},Q_{b},Q_{c}|n \in \bz \}$
be a set of operators satisfying the following commutation relations:
\begin{equation}
\begin{array}{ll}
 \qint{a_{n},a_{m}} = \delta_{n+m,0}
   \fra{\qint{2n}\qint{(k+2)n}}{ n}, &
   \qint{\tilde{a}_{0},Q_{a}}=2(k+2), \hh \\
 \qint{b_{n},b_{m}} = - \delta_{n+m,0}
   \fra{\qint{2n}\qint{2n}}{ n}, &
   \qint{\tilde{b}_{0},Q_{b}}=-4, \hh \\
 \qint{c_{n},c_{m}} = \delta_{n+m,0}
   \fra{\qint{2n}\qint{2n}}{ n}, &
   \qint{\tilde{c}_{0},Q_{c}}=4, \hh
\end{array}
\end{equation}
where
\begin{equation}
\begin{array}{ccc}
 \tilde{a}_{0}=\fra{q-q^{-1}}{ 2\log q} a_{0} , &
 \tilde{b}_{0}=\fra{q-q^{-1}}{ 2\log q} b_{0} , &
 \tilde{c}_{0}=\fra{q-q^{-1}}{ 2\log q} c_{0} ,
\end{array}
\end{equation}
and the other
commutators
vanish.

Let us introduce three free boson fields $a, b,$ and $c$
carrying parameters $L,M,N \in \bz_{>0}$, $\alpha \in \br$.
Define $\bosal{L}{M}{N}{z}{\alpha}$ by
\begin{equation}
\begin{array}{rl}
\bosal{L}{M}{N}{z}{\alpha} = &
- \displaystyle \sum_{n \neq 0}
\frac{\textstyle \qint{Ln} a_{n}}
{\textstyle \qint{Mn}\qint{Nn}} z^{-n}q^{|n|\alpha}
+ \frac{\textstyle L\tilde{a}_{0}}{\textstyle MN} \log z
+ \frac{\textstyle LQ_{a}}{\textstyle MN}.
\end{array}
\end{equation}
We define
$\bosbl{L}{M}{N}{z}{\alpha}, \boscl{L}{M}{N}{z}{\alpha}$
in the same way.
In the case $L=M$ we also write
\begin{equation}
\begin{array}{rl}
\bosa{N}{z}{\alpha} =&
\bosal{L}{L}{N}{z}{\alpha} \\
=&  - \displaystyle \sum_{n \neq 0}
\frac{\textstyle a_{n}}
{\textstyle \qint{Nn}} z^{-n}q^{|n|\alpha}
+ \frac{\textstyle \tilde{a}_{0}}{\textstyle N} \log z
+ \frac{\textstyle Q_{a}}{\textstyle N},
\end{array}
\end{equation}
and likewise for $\bosb{N}{z}{\alpha}, \bosc{N}{z}{\alpha}$.

{\bf 3. $q$-difference operator and the Jackson integral} \qquad
We define a sort of
$q$-difference operator with a parameter $n \in \bz_{>0}$ by
\begin{equation}
\diff{n}{z} f(z) \equiv \frac{\textstyle f(q^{n}z) - f(q^{-n}z)}
{\textstyle (q-q^{-1})z}.
\end{equation}
Then we have the following chain rule:
\begin{equation}
\begin{array}{rl}
\diff{n}{z}\bigl(f(z)g(z)\bigr)= &
(\diff{n}{z} f(z) ) g(q^{n}z) + f(q^{-n}z)(\diff{n}{z} g(z) ) \hh \\
= &(\diff{n}{z} f(z) ) g(q^{-n}z) + f(q^{n}z)(\diff{n}{z} g(z) ) .
\end{array}
\end{equation}

We get formulas for the $q$-difference of boson fields, for example
\begin{equation}
\diff{N}{z} \bosa{N}{z}{\alpha} =
\displaystyle \sum_{n \in \bz} a_{n} z^{-n-1} q^{|n|\alpha}.
\end{equation}
Note that this is independent of $N$.
The operators $\tilde{a}_{0}, \tilde{b}_{0}, \tilde{c}_{0}$
are so normalized that the formula (8) becomes simple.

Now let $p$ be a complex number such that $|p|<1$ and $s \in \bc^{\times}$.
We define the Jackson integral by
\begin{equation}
\int^{s\infty}_{0} f(t) d_{p}t = s(1-p) \displaystyle
\sum^{\infty}_{m=-\infty} f(sp^{m})p^{m} ,
\end{equation}
whenever it is convergent \cite{Ma1}.

If the integrand $f(t)$ is a
total difference of some function $F(t)$:
\begin{equation}
f(t) = \diff{n}{t} F(t),
\end{equation}
then by taking $p=q^{2n}$, we have
\begin{equation}
\int^{s\infty}_{0} f(t) d_{p}t = 0.
\end{equation}

{\bf 4. Wick's Theorem} \qquad
Let
$\{a_{n},b_{n},c_{n}|n \in \bz_{\geq 0} \}$ be annihilation operators,
and $\{a_{n},$ $b_{n},c_{n},Q_{a},Q_{b},Q_{c}|n \in \bz_{< 0} \}$ creation
operators.
We denote by $: \cdots :$ the corresponding normal ordering of operators.
For example,
\begin{equation}
:\exp \left\{ \bosb{2}{z}{\alpha} \right\}: =
\exp \left\{- \displaystyle \sum_{n < 0}
\fra{b_{n}}{\qint{2n}} z^{-n} q^{|n|\ \alpha} \right\}
\exp \left\{- \displaystyle \sum_{n > 0}
\fra{b_{n}}{\qint{2n}} z^{-n} q^{|n|\ \alpha} \right\}
e^{Q_{b}/2} z^{\tilde{b}_{0}/2}.
\end{equation}

The propagator of boson field $a$
reads as follows:
\begin{equation}
\begin{array}{rl}
&\left< \bosal{L}{M}{N}{z}{\alpha} \bosal{L'}{M'}{N'}{w}{\beta} \right> \hh \\
=& - \displaystyle \sum_{n>0} \fra{\qint{Ln} \qint{L'n}
\qint{a_{n} , a_{-n}}}{\qint{Mn} \qint{Nn} \qint{M'n} \qint{N'n}}
\left( \fra{w}{z} \right)^{n} q^{(\alpha + \beta)n} +
\fra{LL'\qint{\tilde{a}_{0} , Q_{a}}}{MNM'N'}\log z \hh \\
=& - \displaystyle \sum_{n>0} \fra{\qint{Ln} \qint{L'n}
\qint{2n} \qint{(k+2)n}}{\qint{Mn} \qint{Nn} \qint{M'n} \qint{N'n}}
\left( \fra{w}{z} \right)^{n} q^{( \alpha + \beta)n} +
\fra{LL'2(k+2)}{MNM'N'}\log z . \hh
\end{array}
\end{equation}
The formal power series in $w/z$ is convergent if $|w/z|<\hskip-2pt<1$.
We introduce the propagators for boson fields $b$ and $c$ in the same manner.
Occasionally one can rewrite
it simply by using the logarithm.
For example,
\begin{equation}
\begin{array}{rl}
\left< \bosb{2}{z}{\alpha} \bosb{2}{w}{\beta} \right> =&
- \log (z-q^{ \alpha+\beta}w) \; , \;\;\;  |z| < |q^{\alpha+\beta}w| .
\end{array}
\end{equation}
Using these propagators, we obtain Wick's Theorem in the following form:

{\bf Proposition 1} ({\it Wick's Theorem})

\begin{equation}
\begin{array}{ll}
{\rm 1)} & \left\{
\begin{array}{rl}
&:\exp \left\{ \bosal{L}{M}{N}{z}{\alpha} \right\}:
:\exp \left\{ \bosal{L'}{M'}{N'}{w}{\beta} \right\}: \hh \\
= &
\exp \left\{ \left< \bosal{L}{M}{N}{z}{\alpha}
\bosal{L'}{M'}{N'}{w}{\beta} \right> \right\} \hh \\
\times &
:\exp \left\{ \bosal{L}{M}{N}{z}{\alpha} +
\bosal{L'}{M'}{N'}{w}{\beta} \right\}: ,
\end{array}  \right. \hh \\
& \\
{\rm 2)} & \left\{
\begin{array}{rl}
&\left[\diff{n}{z} :\exp \left\{ \bosal{L}{M}{N}{z}{\alpha} \right\}: \;\right]
:\exp \left\{ \bosal{L'}{M'}{N'}{w}{\beta} \right\}: \hh \\
= &
\diff{n}{z} \left[
\exp \left\{ \left< \bosal{L}{M}{N}{z}{\alpha}
\bosal{L'}{M'}{N'}{w}{\beta} \right> \right\}  \right. \hh \\
\times &\left.
:\exp \left\{ \bosal{L}{M}{N}{z}{\alpha} +
\bosal{L'}{M'}{N'}{w}{\beta} \right\}: \; \right] ,
\end{array} \right. \hh \\
& \\
{\rm 3)} & \left\{
\begin{array}{rl}
&:\exp \left\{ \bosal{L}{M}{N}{z}{\alpha} \right\}:
\left[ \diff{n}{w} :\exp \left\{ \bosal{L'}{M'}{N'}{w}{\beta} \right\}:
\; \right] \hh \\
= &
\diff{n}{w} \left[
\exp \left\{ \left< \bosal{L}{M}{N}{z}{\alpha}
\bosal{L'}{M'}{N'}{w}{\beta} \right> \right\}  \right. \hh \\
\times & \left.
:\exp \left\{ \bosal{L}{M}{N}{z}{\alpha} +
\bosal{L'}{M'}{N'}{w}{\beta} \right\}: \;\right] . \hh
\end{array} \right.
\end{array}
\end{equation}
{\it There are similar formulas for $b$ and $c$.}

{\bf 5. Current algebra} \qquad
Now we define
the
currents $J^{3}(z), J^{\pm}(z)$
as follows:
\begin{equation}
\begin{array}{rl}
J^{3}(z) =& \diff{k+2}{z} \bosa{k+2}{q^{-2}z}{-1}
+\diff{2}{z} \bosb{2}{q^{-k-2}z}{-{\textstyle \frac{k+2}{2}}} , \hh \\
J^{+}(z) =&
-:\left[\diff{1}{z} \exp \left\{ -\bosc{2}{q^{-k-2}z}{0} \right\} \right]
\times \exp
\left\{ -\bosb{2}{q^{-k-2}z}{1} \right\}: , \hh \\
J^{-}(z) =&
:\left[ \diff{k+2}{z} \exp \left\{
\bosa{k+2}{q^{-2}z}{-{\textstyle \frac{k+2}{2}}}
+\bosb{2}{q^{-k-2}z}{-1} \right. \right. \hh \\
& \hspace{2.5cm} \left. \left.
+\boscl{k+1}{2}{k+2}{q^{-k-2}z}{0} \right\}\right] \hh \\
& \times \exp \left\{-\bosa{k+2}{q^{-2}z}{\frac{k+2}{2}}
+\boscl{1}{2}{k+2}{q^{-k-2}z}{0} \right\}: . \hh
\end{array}
\end{equation}
Define further the
auxiliary fields $\psi(z), \varphi(z)$ as
\begin{equation}
\begin{array}{rl}
\psi(z) =& :\exp \left\{
(q-q^{-1}) \displaystyle
\sum_{n>0} (q^{n}a_{n}+q^{\frac{(k+2)}{2}n}b_{n}) z^{-n}
+ (\tilde{a}_{0}+\tilde{b}_{0}) \log q \right\}: , \hh \\
\varphi(z) =& :\exp \left\{ -(q-q^{-1}) \displaystyle \sum_{n<0}
(q^{3n}a_{n}+q^{\frac{3(k+2)}{2}n}b_{n}) z^{-n}
- (\tilde{a}_{0}+\tilde{b}_{0}) \log q \right\}: . \hh
\end{array}
\end{equation}

By the definition of the boson fields $a,b,c$, and the
$q$-difference operator,
we can recast these fields as
\begin{equation}
\begin{array}{rl}
J^{3}(z) =& \displaystyle \sum_{n \in \bz}
(q^{2n-|n|}a_{n} + q^{(k+2)n-\frac{k+2}{2}|n|}b_{n})z^{-n-1} , \hh \\
J^{+}(z) =& \fra{-1}{(q-q^{-1})z} \left[ :\exp \left\{
-\bosb{2}{q^{-k-2}z}{1}-\bosc{2}{q^{-k-1}z}{0}
\right\}: \right. \hh \\
& \hspace{2.0cm} - \left. :\exp \left\{
-\bosb{2}{q^{-k-2}z}{1}-\bosc{2}{q^{-k-3}z}{0}
\right\}: \right] , \hh \\
J^{-}(z) =& \fra{1}{(q-q^{-1})z} \left[ :\exp \left\{
\bosa{k+2}{q^{k}z}{-\frac{k+2}{2}}
-\bosa{k+2}{q^{-2}z}{\frac{k+2}{2}} \right.\right. \hh \\
&\hspace{3.6cm} \left.\left.+\bosb{2}{z}{-1}
+\bosc{2}{q^{-1}z}{0} \right\}: \right. \hh \\
&\hspace{2.0cm} - \left. :\exp \left\{
\bosa{k+2}{q^{-k-4}z}{-\frac{k+2}{2}}
-\bosa{k+2}{q^{-2}z}{\frac{k+2}{2}} \right.\right. \hh \\
&\hspace{4.0cm} \left.\left.+\bosb{2}{q^{-2k-4}z}{-1}
+\bosc{2}{q^{-2k-3}z}{0} \right\}: \right] , \hh \\
\psi(z) = &:\exp \left\{
\bosa{k+2}{q^{\frac{k}{2}}z}{-\frac{k+2}{2}}
-\bosa{k+2}{q^{-\frac{k}{2}-2}z}{\frac{k+2}{2}} \right. \hh \\
&\hspace{1.3cm} \left. +\bosb{2}{q^{-\frac{k}{2}}z}{-1}
-\bosb{2}{q^{-\frac{k}{2}-2}z}{1} \right\}: , \hh \\
\varphi(z) =& :\exp \left\{
\bosa{k+2}{q^{-\frac{k}{2}-4}z}{-\frac{k+2}{2}}
-\bosa{k+2}{q^{\frac{k}{2}-2}z}{\frac{k+2}{2}} \right. \hh \\
&\hspace{1.3cm} \left. +\bosb{2}{q^{-\frac{3k}{2}-4}z}{-1}
-\bosb{2}{q^{-\frac{3k}{2}-2}z}{1} \right\}: . \hh
\end{array}
\end{equation}
These manageable expressions would be convenient for calculation.
Using Wick's Theorem, we get the following formulas.

{\bf Proposition 2}
{\it The following relations hold in the sense of analytic continuation:}
\begin{equation}
\begin{array}{rl}
\varphi(z) \varphi(w) =& \varphi(w) \varphi(z) , \hh \\
\psi(z) \psi(w) =& \psi(w) \psi(z) , \hh \\
\varphi(z) \psi(w) =&
\fra{(q^{2}z-q^{k}w)(z-q^{-k+2}w)}{(z-q^{k+2}w)(q^{2}z-q^{-k}w)}
\psi(w)\varphi(z), \hh \\
\psi(z) J^{\pm}(w) =& \left(
\fra{q^{2}z-q^{\mp \frac{k}{2}}w}{z-q^{2 \mp \frac{k}{2}}w} \right)^{\pm 1}
J^{\pm}(w) \psi(z) , \hh \\
\varphi(z) J^{\pm}(w) =& \left(
\fra{q^{2}w-q^{\mp \frac{k}{2}}z}{w-q^{2 \mp \frac{k}{2}}z} \right)^{\mp 1}
J^{\pm}(w) \psi(z) , \hh \\
J^{\pm}(z)J^{\pm}(w) =& \fra{q^{\pm2}z-w}{z-q^{\pm2}w}
J^{\pm}(w)J^{\pm}(z) , \hh
\end{array}
\end{equation}
{\it and}
\begin{equation}
J^{+}(z)J^{-}(w) \sim \fra{1}{q-q^{-1}} \left(
\fra{1}{(z-q^{k}w)w} \psi(q^{\frac{k}{2}}w) -
\fra{1}{(z-q^{-k}w)w} \varphi(q^{-\frac{k}{2}}w) \right) .
\end{equation}
{\it The symbol $\sim$ in the last formula means
that both sides of the formula
are equal modulo some fields which are regular at $z=q^{\pm k}w$.}

We consider the mode expansions of these fields as
\begin{equation}
\begin{array}{rlrl}
\displaystyle \sum_{n \in \bz} J^{3}_{n} z^{-n-1} =& J^{3}(z) , &
\displaystyle \sum_{n \in \bz} J^{\pm}_{n} z^{-n-1} =& J^{\pm}(z) , \hh \\
\displaystyle \sum_{n \in \bz} \psi_{n} z^{-n} =& \psi(z) , &
\displaystyle \sum_{n \in \bz} \varphi_{n} z^{-n} =& \varphi(z). \hh
\end{array}
\end{equation}
Putting $K= q^{\tilde{a}_{0}+\tilde{b}_{0}}$
we obtain our main proposition:

{\bf Proposition 3} {\it The operators
$\{J^{3}_{n}|n \in \bz_{\neq 0} \}, \{ J^{\pm}_{n}|
n \in \bz \}$ and $K$ satisfy the following relations.}
\begin{equation}
\begin{array}{l}
\qint{J^{3}_{n},J^{3}_{m}} =
\delta_{n+m,0} \fra{1}{n}\qint{2n} \qint{kn} \, , \,\, n \neq 0 , \hh \\
\qint{J^{3}_{n},K} = 0 , \hh \\
KJ^{\pm}_{n}K^{-1} = q^{\pm2}J^{\pm}_{n} , \hh \\
\qint{J^{3}_{n},J^{\pm}_{m}} =
\pm \fra{1}{n}\qint{2n} q^{\mp\frac{k|n|}{2}}J^{\pm}_{n+m} , \hh \\
J^{\pm}_{n+1}J^{\pm}_{m}-q^{\pm2}J^{\pm}_{m}J^{\pm}_{n+1} =
q^{\pm2}J^{\pm}_{n}J^{\pm}_{m+1}-J^{\pm}_{m+1}J^{\pm}_{n} , \hh \\
\qint{J^{+}_{n},J^{-}_{m}} =
\fra{1}{q-q^{-1}}(q^{\frac{k(n-m)}{2}}\psi_{n+m} -
q^{\frac{k(m-n)}{2}}\varphi_{n+m}). \hh
\end{array}
\end{equation}

These are exactly the relations of the Drinfeld realization of
$U_{q}(\hat{sl_{2}})$ for level $k$ \cite{Dr}.
Thus (18) yields the required bosonization.
One can immediately find that this representation goes to the
Wakimoto representation in the $q \rightarrow 1$ limit.

{\bf 6. screening current} \qquad
Let us define the screening current $J^{S}(z)$ as follows:
\begin{equation}
\begin{array}{rl}
J^{S}(z) =& - :\left[ \diff{1}{z}
\exp \left\{ -\bosc{2}{q^{-k-2}z}{0} \right\} \right]  \hh \\
& \times \exp \left\{ -\bosb{2}{q^{-k-2}z}{-1}
- \bosa{k+2}{q^{-2}z}{-\frac{k+2}{2}} \right\} : . \hh
\end{array}
\end{equation}
Then we get the following proposition.

{\bf Proposition 4}
{\it The commutation relations }
\begin{equation}
\begin{array}{l}
\qint{ J^{3}_{n} , J^{S}(z) } = 0 , \hh \\
\qint{ J^{+}_{n} , J^{S}(z) } = 0 , \hh \\
\qint{ J^{-}_{n} , J^{S}(z) } = \diff{k+2}{z} \left[ z^{n}
:\exp \left\{ -\bosa{k+2}{q^{-2}z}{{\textstyle \frac{k+2}{2}}}
\right\}: \right], \hh
\end{array}
\end{equation}
{\it hold for any $n \in \bz$}.

Therefore, if the Jackson integral of the screening current
\begin{equation}
\int^{s\infty}_{0} J^{S}(t) d_{p}t , \;\;\; p=q^{2(k+2)} \hh
\end{equation}
is convergent, then it commutes with $U_{q}(\hat{sl_{2}})$ exactly.

We note that the operator
$\tilde{b}_{0}+\tilde{c}_{0}$
has the following interesting property:
\begin{equation}
\begin{array}{l}
\qint{ J^{3}(z) , \tilde{b}_{0}+\tilde{c}_{0} } = 0 , \hh \\
\qint{ J^{\pm}(z) , \tilde{b}_{0}+\tilde{c}_{0} } = 0 , \hh \\
\qint{ J^{S}(z) , \tilde{b}_{0}+\tilde{c}_{0} } = 0 , \hh \\
\end{array}
\end{equation}
This operator is useful when we construct the $U_{q}(\hat{sl_{2}})$ modules
in the Fock module of bosons \cite{KQS}.


\vspace{1cm}

{\bf Acknowledgements} \qquad
The author would like to thank H. Awata, T. Eguchi, T. Inami, M. Jimbo,
A. Kato, A. Matsuo, T. Miwa, A. Nakayashiki, K. Ogawa,
Y.-H. Quano, T. Sano, K. Sugiyama, and
Y. Yamada for helpful discussions.
He is grateful to kind hospitality at
YITP and RIMS.


\begin{thebibliography}{99}

\bibitem{DFJMN}Davies B, Foda O, Jimbo M, Miwa T and Nakayashiki A,
Diagonalization of The XXZ Hamiltonian by Vertex Operators,
{\it RIMS preprint}, {\bf 873}, (1992).

\bibitem{FR}Frenkel I B, Reshetikhin N Yu, Quantum Affine Algebras
and Holonomic Difference Equations, {\it Commun. Math. Phys.}
{\bf 146} (1992) 1-60.

\bibitem{IIJMNT}Idzumi M, Iohara K, Jimbo M, Miwa T, Nakashima T, Tokihiro T,
Quantum Affine Symmetry in Vertex Models, {\it RIMS preprint} (1992).

\bibitem{JMMN}Jimbo M, Miki K, Miwa T and Nakayashiki A,
Correlation Functions of the XXZ Model for $\Delta < -1$,
{\it RIMS preprint 877}, (1992).

\bibitem{FJ}Frenkel I B and Jing N H, Vertex Representations of
Quantum Affine Algebras, {\it Proc. Nat'l. Acad. Sci. USA}, {\bf 85},
(1988) 9373-9377.

\bibitem{Dr}Drinfeld V G, A New Realization of Yangians and Quantum
Affine Algebras, {\it Soviet Math. Doklady}, {\bf 36}, (1988) 212-216.

\bibitem{Wa}Wakimoto M, Fock Representations of the Affine Lie
Algebra $A^{(1)}_{1}$, {\it Comm. Math. Phys.}, {\bf 104}, (1986) 605-609.

\bibitem{FMS}Friedan D, Martinec E and Shenker S, Conformal Invariance,
Supersymmetry and String Theory, {\it Nucl. Phys.} {\bf B271}, (1986) 93-165.

\bibitem{Ma1}Matsuo A, Jackson Integrals of Jordan-Pochhammer Type and
Knizhnik-Zamolodchikov Equations, {\it preprint}, (1992).

\bibitem{Ma2}Matsuo A, Quantum Algebra Structure of
Certain Jackson Integrals, {\it preprint}, (1992).

\bibitem{Ma3}Matsuo A, Free Field Representation of Quantum Affine Algebra
$U_{q}(\hat{sl_{2}})$, {\it preprint}, (1992).

\bibitem{Re}Reshetikhin N, Jackson-Type Integrals, Bethe Vectors,
and Solutions to a Difference Analog of the Knizhnik-Zamolodchikov
System, {\it preprint} (1992).

\bibitem{KQS}Kato A, Quano Y-H and Shiraishi J, Free Boson Representation
of $q$-Vertex Operators and Their Correlation Functions, {\it Univ. Tokyo
preprint} {\bf UT-618}, (1992).

\end{thebibliography}
\end{document}